\documentclass[nofootinbib,aps,a4paper,11pt,letterpaper,superscriptaddress,onecolumn,times,eqsecnum]{revtex4}
\usepackage{amsmath}
\usepackage{amsfonts}
\usepackage{graphicx}
\usepackage{color}
\usepackage{booktabs} 
\usepackage{braket}
\usepackage{bm} 
\usepackage{dcolumn}
\usepackage{bm,url}
\usepackage[linktocpage]{hyperref}
\usepackage{subfigure}
\usepackage{amsfonts}
\usepackage[left=2.5cm,right=2.5cm,top=2.5cm,bottom=2.5cm]{geometry}
\usepackage[usenames,dvipsnames,svgnames]{xcolor}  
\usepackage{hyperref}   
\definecolor{oxfordblue}{rgb}{0.0, 0.13, 0.28}
\definecolor{burgundy}{rgb}{0.5, 0.0, 0.13}
\definecolor{darkolivegreen}{rgb}{0.33, 0.42, 0.18}
\definecolor{darkblue}{rgb}{0,0,0.5}
\definecolor{richcarmine}{rgb}{0.84, 0.0, 0.25}
\definecolor{darkblue}{rgb}{0,0,0.5}
\definecolor{bluer}{rgb}{0.00,0.50,0.75}{}
\hypersetup{colorlinks=true, citecolor=cyan, linkcolor=blue,
urlcolor = magenta, filecolor=magenta}

\begin{document}

\newcommand\be{\begin{equation}}
\newcommand\ee{\end{equation}}
\newcommand\bea{\begin{eqnarray}}
\newcommand\eea{\end{eqnarray}}
\newcommand\bseq{\begin{subequations}} 
\newcommand\eseq{\end{subequations}}
\newcommand\bcas{\begin{cases}}
\newcommand\ecas{\end{cases}}
\newcommand{\p}{\partial}
\newcommand{\f}{\frac}

\title{Confronting dark energy in Harada's Conformal Killing Gravity with observational data}
\author{\textbf{Mohsen Khodadi}}
\email{khodadi@kntu.ac.ir}

\affiliation{School of Physics, Damghan University, Damghan 3671641167, Iran}

\affiliation{Center for Theoretical Physics, Khazar University, 41 Mehseti Str., AZ1096 Baku, Azerbaijan}

\author{\textbf{Gaetano Lambiase}}
\email{lambiase@sa.infn.it}
	\affiliation{Dipartimento di Fisica ``E.R Caianiello'', Università degli Studi di Salerno, Via Giovanni Paolo II, 132 - 84084 Fisciano (SA), Italy}
\affiliation{INFN - Gruppo Collegato di Salerno, Via Giovanni Paolo II, 132 - 84084 Fisciano (SA), Italy}

\author{\textbf{Javad. T. Firouzjaee }}
\email{firouzjaee@kntu.ac.ir}
\affiliation{Department of Physics, K. N. Toosi University of Technology,	P. O. Box 15875-4416, Tehran, Iran}


\date{\today}
\begin{abstract}
Based on a comprehensive analysis of recent observational data—a combination of DESI DR1, Planck CMB, and Pantheon+ SN Ia—this study critically evaluates the two dark energy (DE) proposals within Harada's Conformal Killing Gravity (CKG) model. The model at question predicts either a dominant phantom-type effective DE component with EoS \(\omega = -5/3\) or a hybrid scenario combining a cosmological constant (\(\omega = -1\)) with a subdominant \(\omega = -5/3\) fluid (around 5\%) to address the Hubble tension (HT) and late-time acceleration. An analysis based on the Trans-Planckian Censorship Conjecture (TCC) demonstrates that the pure CKG fluid scenario \(\omega = -5/3\) is excluded, whereas the hybrid model remains only marginally compatible.
Our Markov Chain Monte Carlo (MCMC) analysis constrains the effective DE density parameter to \(\Omega_{\text{eff}} = 0.009^{+0.006}_{-0.007}\) (68\% CL), consistent with zero and ruling out the \(\sim 5\%\) contribution required by Harada's CKG. The resulting Hubble expansion history \(H(z)\) and effective EoS \(\omega_{\text{eff}}(z)\) are indistinguishable from those of \(\Lambda\)CDM. Bayesian model comparison via the Akaike Information Criterion (AIC) shows no statistical preference for CKG over \(\Lambda\)CDM (\(\Delta \text{AIC} = +2.6\)), disfavoring the additional complexity of the CKG model. The key output of this study is that both DE proposals in Harada's CKG are ruled out by current cosmological data, and HT remains unresolved.
\\

\end{abstract}

\maketitle
\section{Introduction}

The origin of the present-time accelerating expansion of the universe is commonly attributed to an unknown component known as dark energy (DE), which accounts for around $70\%$ of the energy budget of the universe \cite{Huterer:2017buf}. In the framework of the standard $\Lambda$CDM model, the cosmological constant $\Lambda$ is responsible for DE which in essence is connected with the zero-point vacuum energy density of quantum fields. However, it suffers from the worst fine-tuning problem in theoretical physics, as the value predicted by theoretical considerations is in severe conflict with observational analyses \cite{Bernardo:2022cck}. In well-known Refs. \cite{Weinberg:1988cp,Padmanabhan:2002ji} one can find some theoretical challenges of $\Lambda$CDM model, in details. The standard $\Lambda$CDM model, while successful, faces significant observational tensions, too. Key cosmological parameters derived from its framework are increasingly in conflict with direct measurements. These include: A statistically significant tension in the Hubble constant $H_0$, with Planck CMB data suggesting $67.4\pm0.5$ km s$^{-1}$ Mpc$^{-1}$ \cite{Planck:2018vyg} and the Cepheid-calibrated SnIa method yielding $74.03\pm1.42$ km s$^{-1}$ Mpc$^{-1}$ \cite{Riess:2019cxk}. A discrepancy between large-scale structure observations \cite{Macaulay:2013swa} and the $\Lambda$CDM model-predicted value of $\sigma_8$. A lower matter density parameter from Lyman-
$\alpha$ forest data \cite{BOSS:2014hwf} than what is inferred from the Cosmic Microwave Background (CMB). Several model-independent cosmographic studies such as \cite{Rezaei:2017hon,Lusso:2019akb,Khadka:2020whe} report significant tensions with $\Lambda$CDM predictions, a consistent result that underscores the necessity of investigating viable alternatives to the standard model. These challenges and other considerations have spurred investigations into alternative cosmologies, including modified gravity (e.g., see Refs. \cite{Sotiriou:2008rp,DeFelice:2010aj,Esposito-Farese:2000pbo,Deffayet:2011gz,Fasiello:2013woa,Nojiri:2005vv} (see also rewiev paper \cite{Clifton:2011jh})) and dynamical DE theories \cite{Efstathiou:1999tm,Chevallier:2000qy,Linder:2002et,Jassal:2005qc}. The key motivation for these diverse models is to justify cosmic acceleration within the \textit{''DE paradigm''}.

Recent findings from the Dark Energy Spectroscopic Instrument (DESI)\footnote{A ground-based DE experiment aims to investigate baryon acoustic oscillations (BAO) and structure formation via redshift-space distortions (RSD), utilizing a comprehensive galaxy and quasar redshift survey spanning a wide cosmological volume \cite{DESI:2016fyo}.} suggest that DE may not be constant but instead could evolve and diminish over time \cite{DESI:2024mwx,DESI:2025zgx,DESI:2025fii,DESI:2025ejh,DESI:2025wyn}. This discovery poses a potential challenge to the cosmological constant ($\Lambda$) assumption in the standard $\Lambda$CDM model. Numerous theoretical models have been proposed to explain such time-dependent DE behavior \cite{Lee:2025yvn}. If validated, these observations could profoundly alter our understanding of cosmic expansion and the ultimate fate of the universe.  
Notably, DESI analyses suggest that the DE Equation-of-State (EoS) parameter, $\omega$, may transition from early universe ($\omega<-1$) to late time ( $\omega>-1$) over time—a behavior referred to as quintom-B \footnote{It is worth noting that in conventional scalar field models of DE, such a crossing violates the Null Energy Condition (NEC), requiring non-trivial modifications to the standard $\Lambda$CDM framework. Consequently, various theoretical approaches such as modified gravity \cite{Amendola:1999er,Saridakis:2012jy,Yang:2025kgc,Yang:2025mws}, and interacting DE \cite{Giare:2024smz,Li:2024qso,Pan:2025qwy} have been proposed to reconcile the quintom scenario with NEC preservation.} \cite{Feng:2004ad,Cai:2009zp}. 

An exciting point is that the observed weakening of the DE-EoS aligns with certain expectations from quantum gravity (see, e.g., Refs. \cite{Payeur:2024kyy,Arjona:2024dsr,Bhattacharya:2024kxp,Brandenberger:2025hof,Anchordoqui:2025fgz}). A key example is the Trans-Planckian Censorship Conjecture (TCC) \cite{Bedroya:2019snp}, which asserts that trans-Planckian quantum fluctuations must not become classical and observable on cosmological scales. If such fluctuations were to propagate beyond the Planck scale, it would imply that unknown trans-Planckian physics could influence low-energy phenomena—undermining the validity of our effective field theories in describing cosmic evolution. Crucially, this would challenge their consistency with a UV-complete theory of quantum gravity \cite{Martin:2000xs,Brandenberger:2012aj}. More exactly,
TCC forms part of the broader Swampland Conjectures framework, initially introduced in Ref. \cite{Vafa:2005ui}. These conjectures establish necessary conditions for effective field theories (EFTs) to be consistent with quantum gravity, effectively distinguishing viable theories from those relegated to the swampland of incompatible models. While extensive prior research has utilized the TCC to establish constraints on inflation scenarios and primordial black hole formation in the early universe, this framework also proves equally powerful in analyzing late-time cosmological evolution. In particular, the TCC provides meaningful constraints on various quintessence scalar-field models \cite{Heisenberg:2018yae,Agrawal:2018own,Cicoli:2020cfj} (see also \cite{Anchordoqui:2025fgz}) that describe DE dynamics in the contemporary universe. The TCC imposes a fundamental constraint on cosmological models within the quantum gravity EFT: any scenario predicting eternal future acceleration is necessarily excluded. This prohibition arises because perpetual expansion would redshift sub-Planckian quantum fluctuations to super-horizon scales, forcing their classicalization--a direct violation of the conjecture's core principle. Recently, in Ref. \cite{Li:2025cxn}, the authors presented a comprehensive investigation of the quintom-B behavior indicated by DESI results (in particular Data Release 2 (DR2)). Their analysis rigorously establishes TCC constraints across different parameterizations of the DE-EoS $\omega(a)$, along with the application of TCC to modified gravity theories, leveraging recent observational datasets to derive novel constraints.

The differential form of TCC imposes fundamental constraints on cosmological evolution, requiring that the EoS parameter asymptotically satisfy $\omega>-1/3$ as $t\rightarrow \infty$ \cite{Li:2025cxn}. This condition necessarily results in a late-time transition to deceleration, thereby naturally bounding the duration of cosmic acceleration. Consequently, the TCC predicts an eventual cessation of universal expansionary acceleration. While our universe has undergone accelerated expansion in its recent past, the TCC necessitates an eventual transition to deceleration in the late-time future. A constant DE-EoS with $\omega>-1/3$ at all epochs would forbid any period of cosmic acceleration. That means the pressure is not negative enough to cause cosmic acceleration. Consequently, the TCC predicts an eventual cessation of universal expansionary acceleration. A constant DE-EoS with \(\omega > -1/3\) at all epochs would forbid any period of cosmic acceleration, as the pressure would not be negative enough. Conversely, a constant EoS with \(\omega < -1/3\) (like \(\Lambda\)CDM or a phantom \(\omega\)CDM model) would cause eternal acceleration, violating the TCC. Therefore, the conjunction of observed current acceleration and the TCC future bound requires a time-varying DE-EoS that crosses the \(\omega = -1/3\) threshold, effectively ruling out both the \(\Lambda\)CDM model and any constant-\(\omega\) model that matches current observations as complete descriptions of cosmic evolution \cite{Li:2025cxn}.
As a result, to model dynamical DE scenarios, parameterized forms of the DE-EoS are commonly employed. A widely used approach is the $\omega_0\omega_a$CDM parameterization, which describes the evolution of $\omega(a)$ with CPL (Chevallier-Polarski-Linder) parameterization $\omega(a) = \omega_0 + \omega_a(1 - a)$ \cite{Chevallier:2000qy,Linder:2002et} through two free parameters. As evident from the parameterization, when $a \rightarrow 1$ (present epoch), the EoS approaches $\omega(a) \rightarrow \omega_0$, representing the current value of the DE-EoS. Conversely, in the early universe ($a \rightarrow 0$), the EoS evolves to $\omega(a) \rightarrow \omega_0+\omega_a$, capturing its asymptotic past behavior.

As before noted, the extended theories of gravity, which generalize Einstein's framework, have emerged as leading candidates to explain the universe's current accelerated expansion. By generating cosmological solutions that go beyond the $\Lambda$CDM paradigm, these modified gravity theories offer compelling alternatives to a cosmological constant while remaining consistent with observational constraints. In a significant theoretical advance, Junpei Harada \cite{Harada:2023rqw,Harada:2023afu} proposed a novel modified gravity theory-Conformal Killing Gravity (CKG)\footnote{Following its release, the gravitational physics of this theory have been investigated and evaluated in recent years, see  \cite{Barnes:2023uru,Barnes:2023qfi,Clement:2024xmr,Mantica:2024sdy,Alshal:2024tcr,Mantica:2024vyg,Chen:2025fse,Ghaffari:2025qmv}. In this regards, pp-waves (plane-fronted waves with parallel rays) of this modified gravity studied in Refs. \cite{Barnes:2024gqy, Gurses:2024ltc}. It is recommend to see Refs. \cite{Mantica:2024mun,Mantica:2023stl,Capozziello:2025kws}, too.}—constructed from third-order derivatives of the metric tensor. This framework naturally generates two distinct DE components: A cosmological constant arising as an integration constant \cite{Mantica:2024mun}, and an effective dark fluid with a characteristic EoS parameter $\omega=-5/3$. This unique dual-component structure distinguishes CKG from conventional modified gravity theories and provides new avenues to address present cosmic acceleration.
In other words, unlike traditional modified gravity theories, CKG's third-order derivative structure induces a DE sector with a fixed-parameter fluid ($\omega=-5/3$) alongside an emergent cosmological constant, providing a new mechanism for late-time acceleration. It is interesting to note that the authors in Ref. \cite{Mantica:2023stl} have published the same result and clarified that the CKG is actually an extended version of Einstein's equation obtained by including an arbitrary killing tensor. 

This work aims to critically evaluate the claim made by Harada's CKG model regarding the universe's late-time acceleration by confronting it with observational data. Our analysis aims to provide novel insights into the validity of proposals made about the DE in this model. This manuscript is structured as follows. Following a brief review of the key discussions in \cite{Harada:2023afu} (Sec. \ref{sec:effective}), we examine Harada's CKG in conjunction with the fundamental TCC criterion in Sec. \ref{tcc}. In Sec. \ref{mcmc}, we employ the Markov Chain Monte Carlo (MCMC) method to numerically assess the model's viability against recent observational data. Finally, we present a discussion and a summary of our results in Sec. \ref{co}.

\section{effective DE \label{sec:effective}}
Recently, Junpei Harada in Refs.~\cite{Harada:2023rqw,Harada:2023afu} proposed a new gravitational theory, referred to as CKG, by demanding the simultaneous satisfaction of three key theoretical criteria for theories extending beyond general relativity \footnote{The primary motivation for this new gravitational theory stems from the fact that existing extended frameworks of gravity—including general relativity—do not simultaneously meet all three of these criteria. The theoretical appeal of this model is substantially strengthened by the recent establishment of an underlying action principle \cite{Feng:2024rnh}.}:  
i- The cosmological constant must emerge as an integration constant rather than being introduced ad hoc. ii- The energy conservation law should be derived from the field equations, not imposed as an independent assumption. iii- Conformally flat metrics should not necessarily be treated as vacuum solutions by default.  

This framework leads to a generalized form of the Friedmann equation as follows, 
\begin{equation}
	2\left(\frac{\dot{a}(t)}{a(t)}\right)^2 - \frac{\ddot{a}(t)}{a(t)} = \frac{4\pi G}{3}(5\rho(t) + 3p(t)) - \frac{2k}{a^2(t)} + \frac{\Lambda}{3},
	\label{eq:Harada_Friedmann}
\end{equation}
offering a novel approach to gravitational dynamics.  If we model the universe as consisting solely of matter (m) and radiation (r), without introducing DE, Eq.~\eqref{eq:Harada_Friedmann} reduces to
\begin{equation}
	2\left(\frac{\dot{a}}{a}\right)^2 - \frac{\ddot{a}}{a} = \frac{4\pi G}{3}(5\rho_{\rm m} + 6\rho_{\rm r}) - \frac{2k}{a^2} + \frac{\Lambda}{3},
	\label{eq:Harada_Friedmann3}
\end{equation}
since the energy density $\rho$ and pressure $p$ in Eq.~\eqref{eq:Harada_Friedmann} are given by $\rho=\rho_{\rm m} + \rho_{\rm r}$ and $p=\rho_{\rm r}/3$, respectively. Here, the energy density $\rho_{\rm m}$ and $\rho_{\rm r}$ as functions of $a$ can be derived from the conservation law $\nabla_\mu T^\mu{}_{\nu}=0$, 
\begin{eqnarray}
\nabla_\mu T^\mu{}_0 =0\Longrightarrow \dot{\rho}+3\frac{\dot{a}}{a}(\rho + p)=0.
	\label{eq:conservation}
\end{eqnarray}
‌By setting $p=0$, and $p=\rho_{\rm r}/3$ in Eq.~\eqref{eq:conservation} gives respectively the following expressions for matter and radiation
\begin{subequations}
	\begin{eqnarray}
		\rho_{\rm m} (t) &= \rho_{\rm m,0}\left(\frac{a(t)}{a_0}\right)^{-3},\\
		\rho_{\rm r} (t) &= \rho_{\rm r,0}\left(\frac{a(t)}{a_0}\right)^{-4},
	\end{eqnarray}
\end{subequations}
where $\rho_{\rm m,0}$ and $\rho_{\rm r,0}$ represent the density for matter (m) and radiation (r) at the present time, respectively. The $a_0$ denotes the scale factor at the present time.

The left-hand side of Eq.~\eqref{eq:Harada_Friedmann3} can be expressed as 
\begin{eqnarray}
	2\left(\frac{\dot{a}}{a}\right)^2 - \frac{\ddot{a}}{a} = H^2-\dot{H}.
	\label{eq:lhs_GFE3}
\end{eqnarray}
where by inserting it into Eq.~\eqref{eq:Harada_Friedmann3} and dividing by $H_0^2$, the Eq.~\eqref{eq:Harada_Friedmann3} can be expressed as 
\begin{eqnarray}
	&& \frac{H^2 - \dot{H}}{H_0^2}  \nonumber\\
	=
	&& \frac{5}{2} \Omega_{\rm m} \left(\frac{a}{a_0}\right)^{-3}
	+3 \Omega_{\rm r} \left(\frac{a}{a_0}\right)^{-4}
	+2 \Omega_k \left(\frac{a}{a_0}\right)^{-2} 		
	+\Omega_\Lambda.\nonumber\\
	\label{eq:Harada_Friedmann4}
\end{eqnarray}
Here, the density parameters $\Omega$'s are defined as follows: 
\begin{eqnarray}
	\Omega_{\rm m} \equiv \frac{\rho_{\rm m,0}}{\rho_{\rm c}},\
	\Omega_{\rm r}  \equiv \frac{\rho_{\rm r,0}}{\rho_{\rm c}},\
	\Omega_k \equiv - \frac{k}{a_0^2 H_0^2},\
	\Omega_\Lambda \equiv \frac{\Lambda}{3H_0^2},\qquad
\end{eqnarray}
and the critical density is defined as $\rho_{\rm c} \equiv 3H_0^2/8\pi G$.
Equation~\eqref{eq:Harada_Friedmann4} is derived without explicitly introducing a DE component into the universe's energy content (i.e., the energy-momentum tensor contains only matter and radiation). The term \(\Omega_{\Lambda}\) originates from an integration constant in the CKG field equations, effectively mimicking a cosmological constant but arising from the geometry of the theory.

The density parameters mentioned above do not obey the standard Friedmann constraint, $\Omega_{\rm m} + \Omega_{\rm r} + \Omega_k + \Omega_\Lambda=1$, which describes the present-day universe in conventional cosmology. Rather, they adhere to the modified relation
\begin{equation}
	2+q_0=\frac{5}{2}\Omega_{\rm m} + 3\Omega_{\rm r} + 2 \Omega_k + \Omega_\Lambda.
	\label{eq:q0_Omega1}
\end{equation}
Here, the deceleration parameter $q$ is defined by
\begin{eqnarray}
	q\equiv - \frac{\ddot{a}a}{\dot{a}^2} = -\frac{\ddot{a}}{aH^2}=-\frac{\dot{H}}{H^2}-1,
	\label{def:deceleration}
\end{eqnarray}
and $q_0$ represents its present value so that if $q_0>0$ the current universe has decelerating expansion while for $q_0<0$ accelerating.
Equation~\eqref{eq:q0_Omega1} indicates that $q_0$ can have a negative value if the following relation is satisfied:
\begin{eqnarray}
	\frac{5}{2}\Omega_{\rm m} + 3\Omega_{\rm r} + 2 \Omega_k + \Omega_\Lambda < 2.
\end{eqnarray}
Contrary to what is expected from general relativity, $q_0$ can become negative even when $\Omega_\Lambda=0$ \footnote{In general relativity the counterpart of Eq. \eqref{eq:q0_Omega1} takes the form $q_0=\frac{1}{2}\Omega_{\rm m}+\Omega_{\rm r}-\Omega_{\rm \Lambda}$ so that if $\Omega_{\rm \Lambda}=0$, then $q_0>0$. This indicates a decelerating expansion of the current universe—in stark contrast to observational expectations.}. For instance, in the case of a matter-dominated universe ($\Omega_{\rm r}=\Omega_k=\Omega_\Lambda=0$), $q_0$ is negative if $\Omega_{\rm m} <0.8$ which is consistent with a previous study~\cite{Harada:2023rqw}). Consequently, within the cosmological framework described by Eq.~\eqref{eq:Harada_Friedmann4}, the present-day expansion of the universe can be accelerating ($q_0<0$), all without the necessity of negative pressure or a cosmological constant.

An explicit solution for the scale factor $a(t)$ was obtained~\cite{Harada:2023rqw} by assuming a matter-dominated universe with $\Omega_{\rm r}=\Omega_k=\Omega_\Lambda=0$. This solution describes the transition from decelerated to accelerated expansion. In the subsequent discussion, we will clearly explain the mechanisms that facilitate this acceleration.

Equation~\eqref{eq:Harada_Friedmann4} includes a time derivative term, $\dot{H}$, which makes it into a differential equation for the Hubble parameter. When we solve Eq.~\eqref{eq:Harada_Friedmann4} for $H$, we obtain
\begin{eqnarray}
	\left(\frac{H}{H_0}\right)^2 
	=\Omega_{\rm m}\left(\frac{a}{a_0}\right)^{-3} 
	+ \Omega_{\rm r}\left(\frac{a}{a_0}\right)^{-4}
	+ \Omega_k\left(\frac{a}{a_0}\right)^{-2} + \Omega_\Lambda
	+ \Omega_{\rm eff}\left(\frac{a}{a_0}\right)^2.\qquad
	\label{eq:Hubble}
\end{eqnarray}
with an additional effective density parameter compared to $\Lambda$CDM model
\begin{eqnarray}
	\Omega_{\rm eff} \equiv 1 - \Omega_{\rm m} - \Omega_{\rm r} - \Omega_k - \Omega_\Lambda.
	\label{eq:five_parameters}
\end{eqnarray}
Notably, Eq.~\eqref{eq:Hubble} is an exact solution to Eq.~\eqref{eq:Harada_Friedmann4} that in terms of the redshift parameter, re-express as follows
\begin{align}
H(z)_{\text{CKG}} =  \left [\Omega_{\rm m} (1+z)^3+\Omega_{\rm r} (1+z)^4 +\Omega_{\rm k}(1+z)^2 +\Omega_\Lambda+\Omega_{\rm eff} (1+z)^{-2} \right ]^{1/2} H_0 \label{HARADAH}
\end{align}


Equation~\eqref{eq:Hubble} generalizes the standard Friedmann equation by adding a term, $\Omega_{\rm eff}(a/a_0)^2$. As is evident, the additional term is characterized by the exponent 2, a value that uniquely arises as a consequence of Eq.~\eqref{eq:Harada_Friedmann4}. Given that the general solution of  Eq.~\eqref{eq:conservation}, takes the form $\rho \propto \left(\frac{a}{a_0}\right)^{-3(1+\omega)}$, then one can deduce that $2=-3(1+\omega)$. It means that the term $\Omega_{\rm eff}(a/a_0)^2$ effectively takes on the role of phantom-type DE with $\omega=-5/3$.

Overall, Harada in 
\cite{Harada:2023afu}, by ignoring the contributions of radiation and curvature ($\Omega_r=0=\Omega_k$), shows that if the effective DE density parameter $\Omega_{\rm eff}= 1-\Omega_m -\Omega_\Lambda $ is present in moderate amounts, CKG has the potential to resolve the Hubble Tension (HT). This statement is valid if $\Omega_{\rm eff}$ is dominant in the total energy budget with $\Lambda=0$.

Consequently, Harada's CKG framework is posited to resolve the well-known Hubble Tension (HT) and, simultaneously, explain the present accelerated expansion of the universe. More precisely, the CKG can explain in the light of two equivalent descriptions, the current accelerating expansion of the universe. First, in the absence of the concept of DE \cite{Harada:2023rqw}. Second, in the presence of a small amount (around $5\%$) of the effective DE concept with EoS parameter, $\omega=-5/3$, together with a large amount of the cosmological constant (around $70\%$) with $\omega=-1$ \cite{Harada:2023afu}. The key point about the effective DE in CKG is that in the absence of it, the HT issue cannot be solved.
Although Harada stresses the equivalence of the two descriptions, he argues that for practical applications, the latter is convenient because it essentially modifies the dynamic equations of the universe on a large scale, by including an extra term $\Omega_{eff}(a/a_0)^2$ in the Friedman equations \cite{Harada:2023afu}. In other words, the advantage of the second approach is that it allows us to test CKG within the DE paradigm observationally—both in the context of $\Lambda$CDM and beyond it.

Briefly, in the second interpretation of the Harada model \cite{Harada:2023afu}, there are the following two proposals to explain the current accelerated expansion of the universe and to resolve the HT.
\begin{enumerate}
	 	\item In presence of both components with EoS parameters $\omega=-1$, and $\omega=-5/3$ (with such a distributions for the density parameters: $\Omega_m\sim25 \%$, $\Omega_{\Lambda}\sim70\%$, and $\Omega_{eff }\sim5\%$).
	 \item	Without a cosmological constant ($\Omega_{\Lambda}=0$), provided that the effective DE component is dominant ($\Omega_{eff }\sim 70\%$).
\end{enumerate}

\section{{Evaluating two DE proposals with the TCC Criterion}}\label{tcc}
The standard TCC criterion in its differential form is a condition on the acceleration of the scale factor
\begin{equation}
\frac{d^2 a}{dt^2} < 0 \quad \text{or} \quad \ddot{a} < 0 \label{TCC} 
\end{equation}
This ensures that the expansion does not accelerate too quickly, preventing trans-Planckian modes from crossing the horizon.
We start with the key equation of CKG (Eq.(\ref{eq:Harada_Friedmann}) for a flat universe, $k=0$)
\begin{equation}
2\left(\frac{\dot{a}}{a}\right)^{2} -\frac{\ddot{a}}{a} = \frac{4\pi G}{3}(5\rho + 3p) + \frac{\Lambda}{3} \label{1}
\end{equation}
Our goal is to express $\ddot{a} < 0$ in terms of the energy content ($\rho, p$) using the CKG equation. We begin by rewriting Eq. (\ref{1}) to isolate $\ddot{a}$
\begin{equation}
2H^2 - \frac{\ddot{a}}{a} = \frac{4\pi G}{3}(5\rho + 3p) + \frac{\Lambda}{3}
\end{equation}
Solving for $\ddot{a}$ yields
\begin{equation}
\ddot{a} = -a \left[ \frac{4\pi G}{3}(5\rho + 3p) + \frac{\Lambda}{3} - 2H^2 \right] \label{2}
\end{equation}
Since the scale factor $a > 0$, the sign of $\ddot{a}$ is determined by the expression in the brackets. Therefore, the TCC condition $\ddot{a} < 0$ implies
\begin{equation}
\frac{4\pi G}{3}(5\rho + 3p) + \frac{\Lambda}{3} - 2H^2 > 0 \label{3}
\end{equation}
For the CKG effective fluid ($\omega = -5/3$), the first term vanishes, meaning that
\begin{equation}
	\frac{\Lambda}{3} - 2H^2 > 0
\end{equation}
Given that if DE dominates, $H^2 \approx \frac{8\pi G}{3}\rho_\text{de} + \frac{\Lambda}{3}$, so in case of $\Lambda=0$ and pure CKG fluid dominates (second proposal of DE in the underlying model), then $H^2 \approx \frac{8\pi G}{3}\rho_{\text{eff}}$, and the condition becomes $- 2H^2 > 0$, which is never true since $H^2 > 0$. Therefore, a universe dominated by the CKG effective fluid ($w=-5/3$) always violates the TCC. This theoretical result provides an independent reason to disfavor the dominant CKG fluid scenario.

Let us consider the first DE proposal, i.e., the case where both $\Lambda$ and $\Omega_{\text{eff}}$ are at play. We start from the general TCC condition (\ref{3}). Now, we consider the universe's content. For the late-time universe, the total energy density $\rho$ and pressure $p$ have contributions from matter ($\rho_m, p_m=0$), radiation (negligible), and the two DE components:  $\rho_\Lambda, \quad p_\Lambda = -\rho_\Lambda, \quad w = -1$, and  $\rho_{\text{eff}}, \quad p_{\text{eff}} = -\frac{5}{3}\rho_{\text{eff}}, \quad w = -5/3$. Therefore, the total density and pressure are: $\rho = \rho_m + \rho_\Lambda + \rho_{\text{eff}}$, and $
p = p_\Lambda + p_{\text{eff}} = -\rho_\Lambda - \frac{5}{3}\rho_{\text{eff}}
$ which by inserting these into the TCC condition, we come to
\begin{equation}
\frac{4\pi G}{3}(5\rho_m + 2\rho_\Lambda) + \frac{\Lambda}{3} - 2H^2 > 0
\end{equation}
We can now express this in terms of density parameters. Recall the definitions:
$\Omega_i = \frac{8\pi G}{3H^2}\rho_i$, and $\Omega_\Lambda = \frac{\Lambda}{3H^2}$. Therefore: $\frac{4\pi G}{3}\rho_m = \frac{4\pi G}{3} \cdot \frac{3H^2}{8\pi G} \Omega_m = \frac{H^2}{2} \Omega_m$, $\frac{4\pi G}{3}\rho_\Lambda = \frac{4\pi G}{3} \cdot \frac{3H^2}{8\pi G} \Omega_\Lambda = \frac{H^2}{2} \Omega_\Lambda$, and $\frac{\Lambda}{3} = H^2 \Omega_\Lambda$.
Substitute these in TCC condition above, after a bit manipulation, we have 
\begin{equation}\label{in}
\frac{5}{2}\Omega_m + 2\Omega_\Lambda - 2 > 0
\end{equation}
This is the final TCC condition for the CKG model with both $\Lambda$ and $\Omega_{\text{eff}}$. Now, let us analyze the inequality (\ref{in}). 

\begin{figure}[ht!]
	\centering
	\includegraphics[width=0.4\columnwidth]{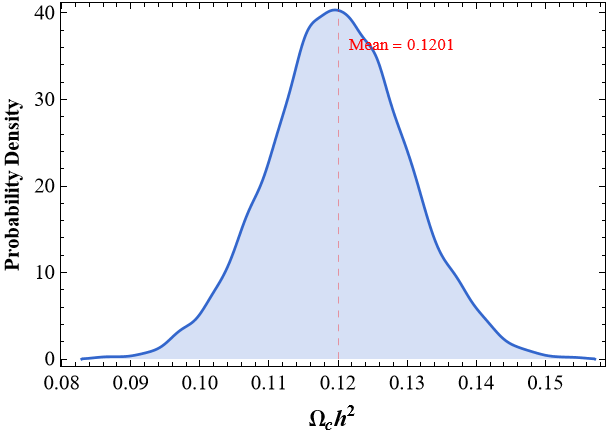}~~~
	\includegraphics[width=0.4\columnwidth]{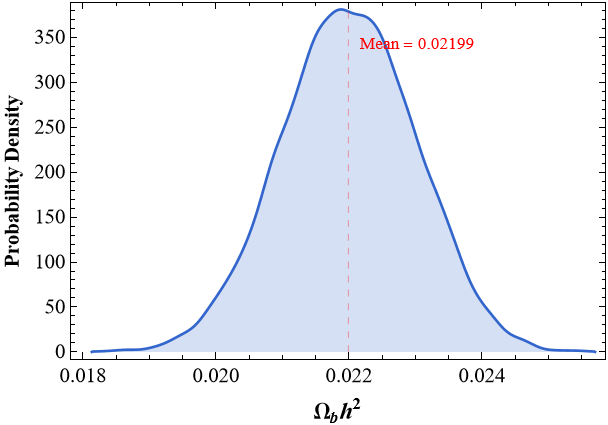}\\
	\includegraphics[width=0.4\columnwidth]{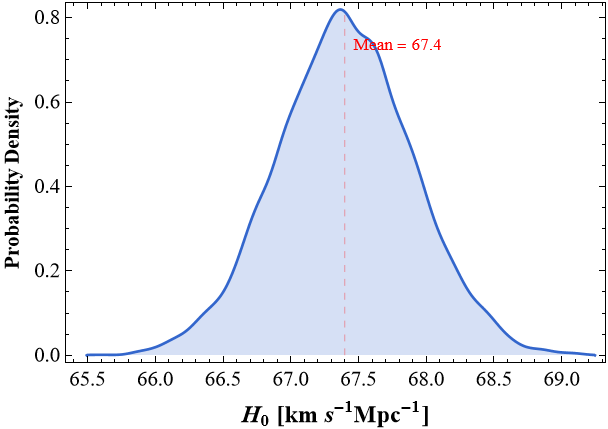}~~~
	\includegraphics[width=0.4\columnwidth]{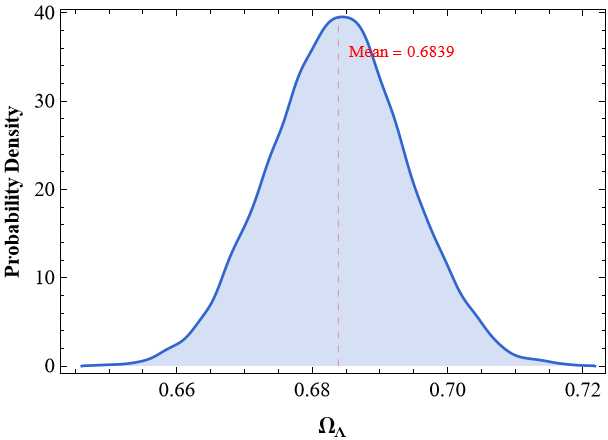}
	\includegraphics[width=0.4\columnwidth]{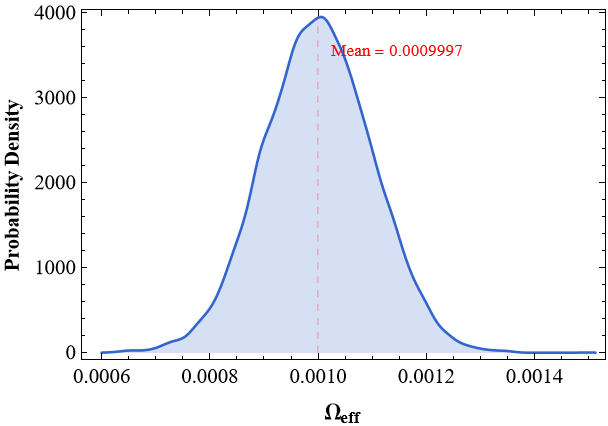}~~~
	\includegraphics[width=0.4\columnwidth]{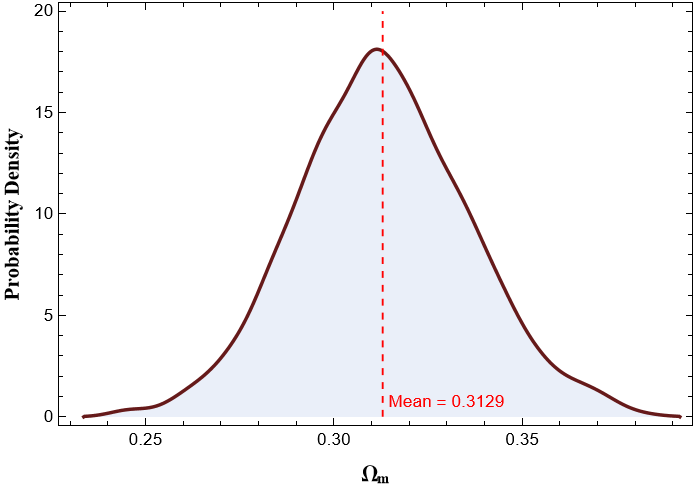}
	\caption{\textit{1D marginalized posterior distributions, for the key parameters of the Harada's CKG model.}} 
	\label{rc1}
\end{figure}
\begin{figure}[ht!]
	\centering
	\includegraphics[width=0.48\columnwidth]{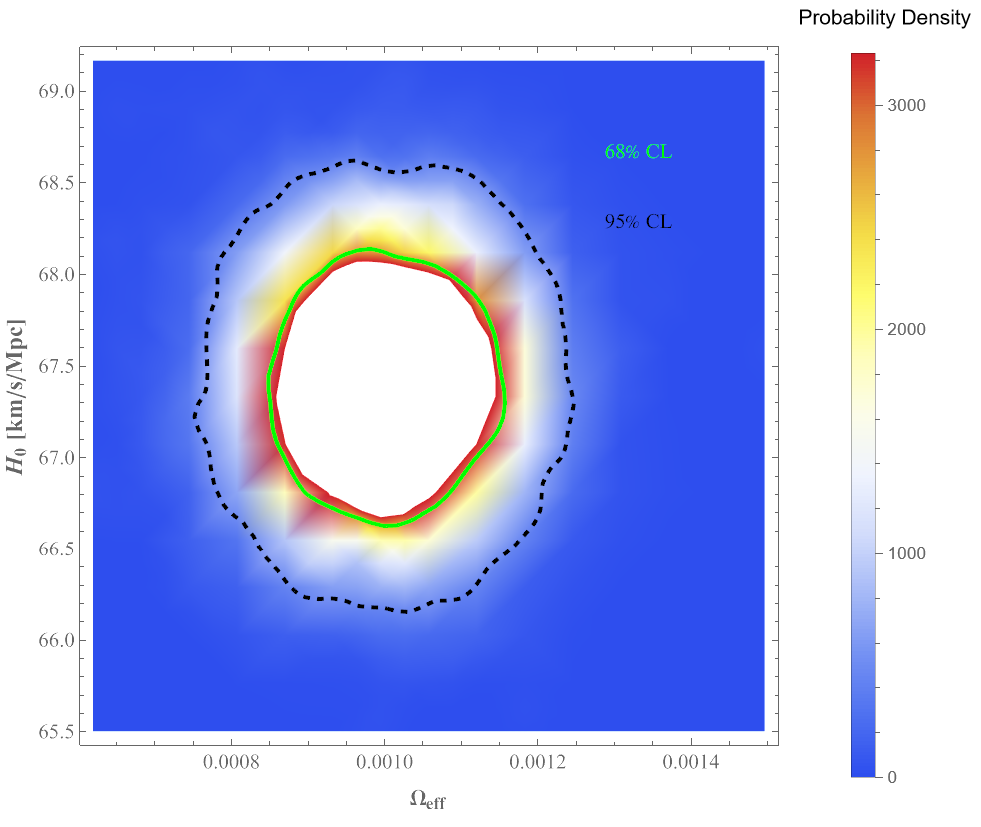}~~
	\includegraphics[width=0.48\columnwidth]{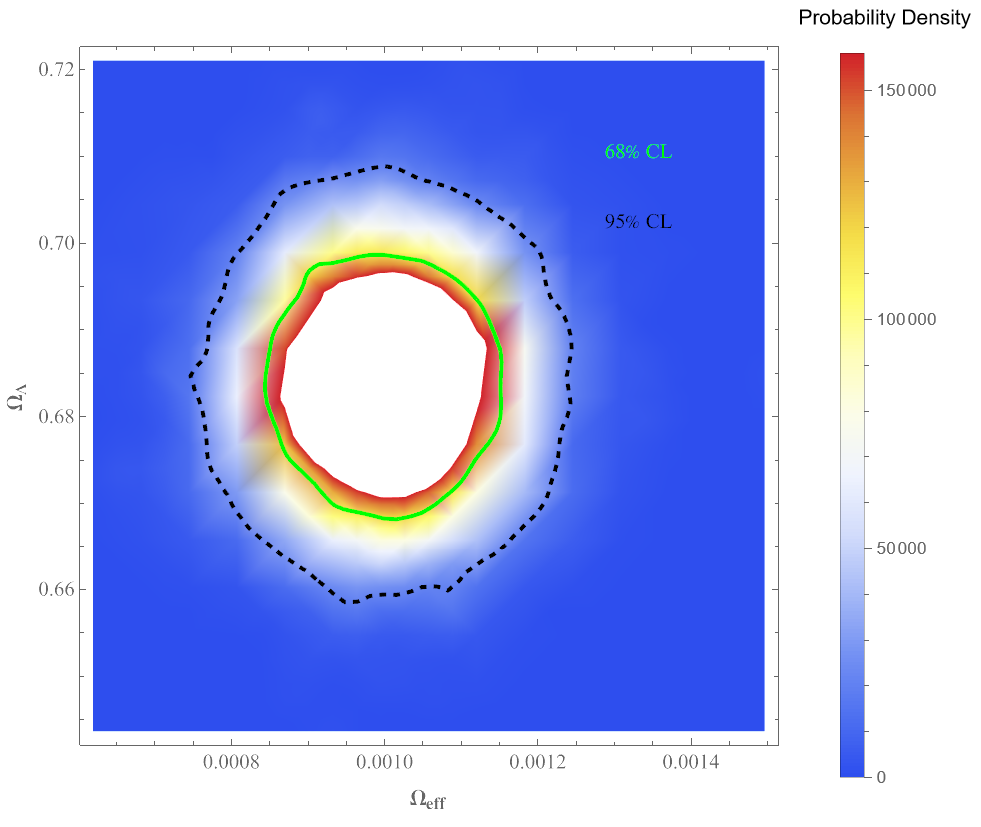}
	\includegraphics[width=0.48\columnwidth]{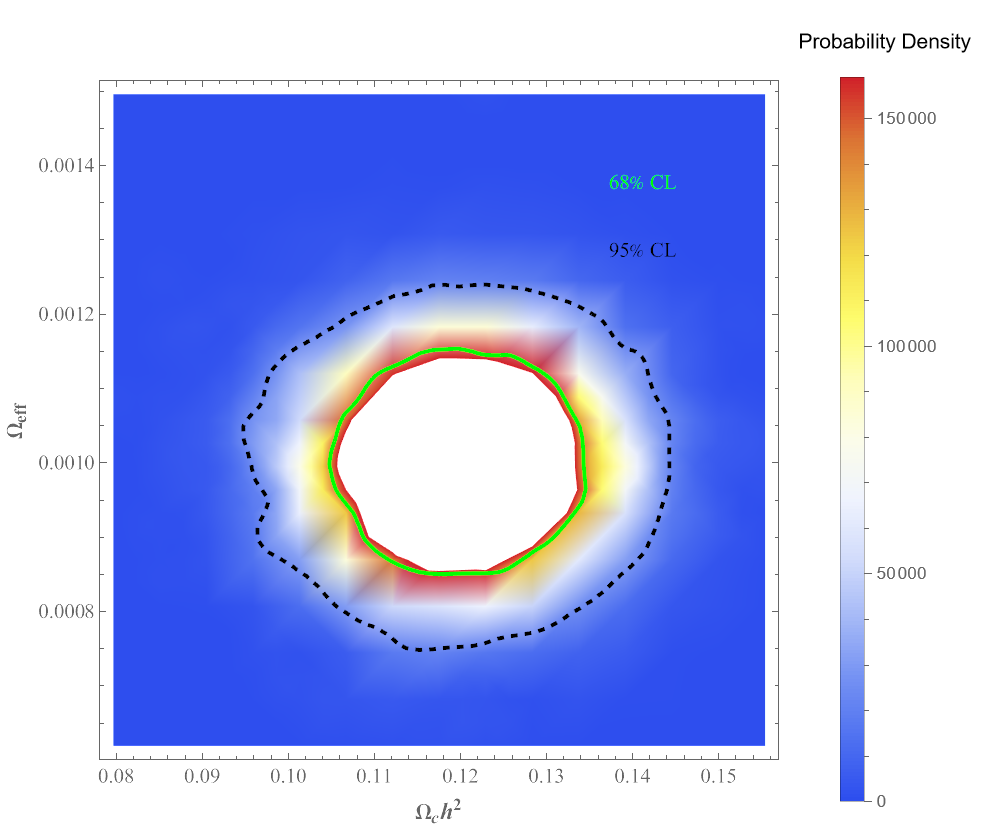}~~
	\includegraphics[width=0.48\columnwidth]{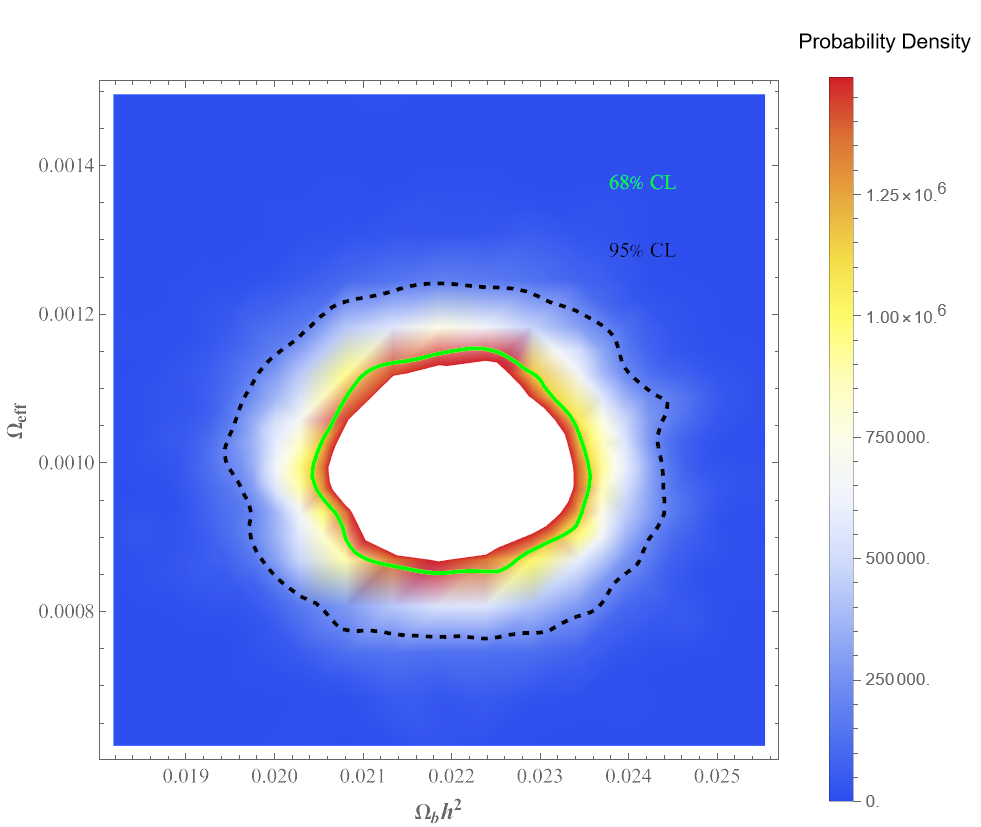}
	\includegraphics[width=0.48\columnwidth]{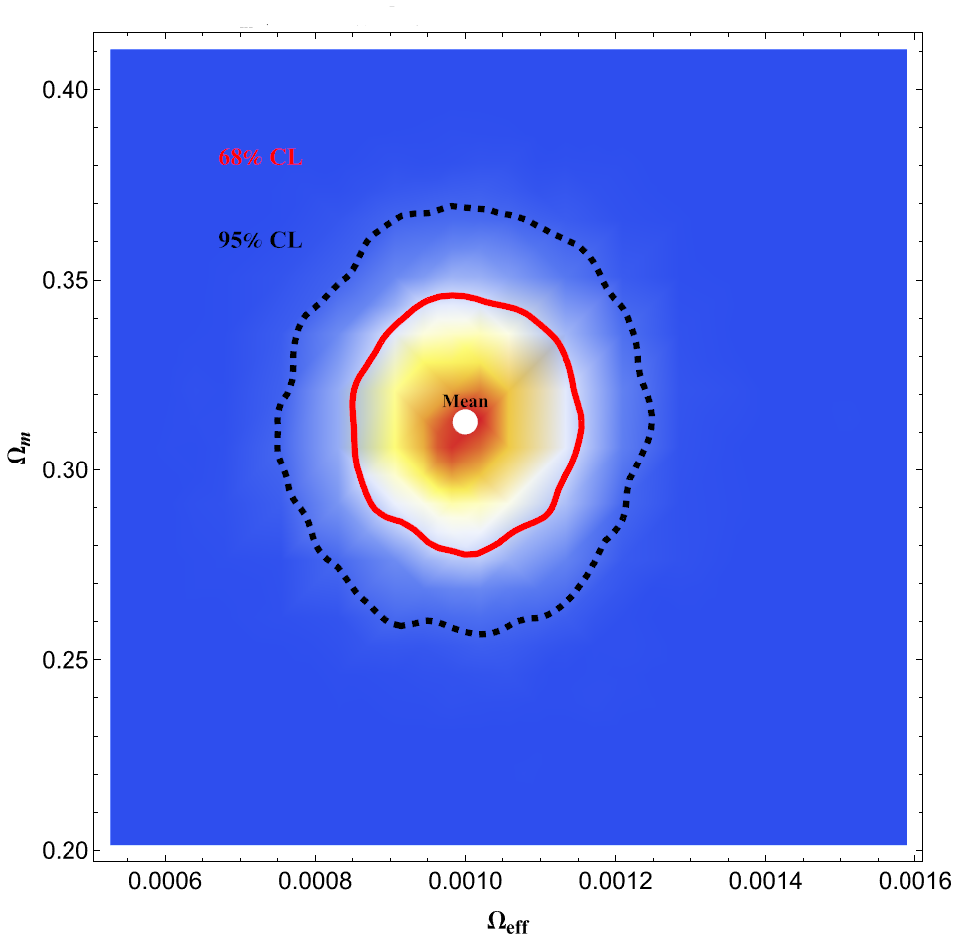}~~
	\caption{\textit{2D posterior distributions within 68\%, and 95\%, for the key parameters of the Harada's CKG model.}} 
	\label{rc2}
\end{figure}

In a $\Lambda$CDM universe ($\Omega_{\text{eff}}=0$) the condition becomes $\frac{5}{2}\Omega_m + 2\Omega_\Lambda > 2$. For the standard values $\Omega_m \approx 0.3$, $\Omega_\Lambda \approx 0.7$ leads to $2.15 > 2$, meaning that 
$\Lambda$CDM is compatible with the TCC. In the CKG proposal ($\Omega_m \sim 0.25$, $\Omega_\Lambda \sim 0.7$, $\Omega_{\text{eff}} \sim 0.05$) \cite{Harada:2023afu}, the TCC condition does not depend on $\Omega_{\text{eff}}$ due to the cancellation which leads to $2.025 > 2$ is (barely) satisfied. As a result, his specific proposal is not ruled out by the TCC. The specific EoS of the CKG fluid ($w = -5/3$) causes its energy density to cancel out of the fundamental TCC condition. Therefore, the TCC cannot rule out the two-component proposal on theoretical grounds; it only constrains the other components ($\Omega_m$ and $\Omega_\Lambda$).

\begin{figure}[ht!]
	\centering
	\includegraphics[width=0.6	\columnwidth]{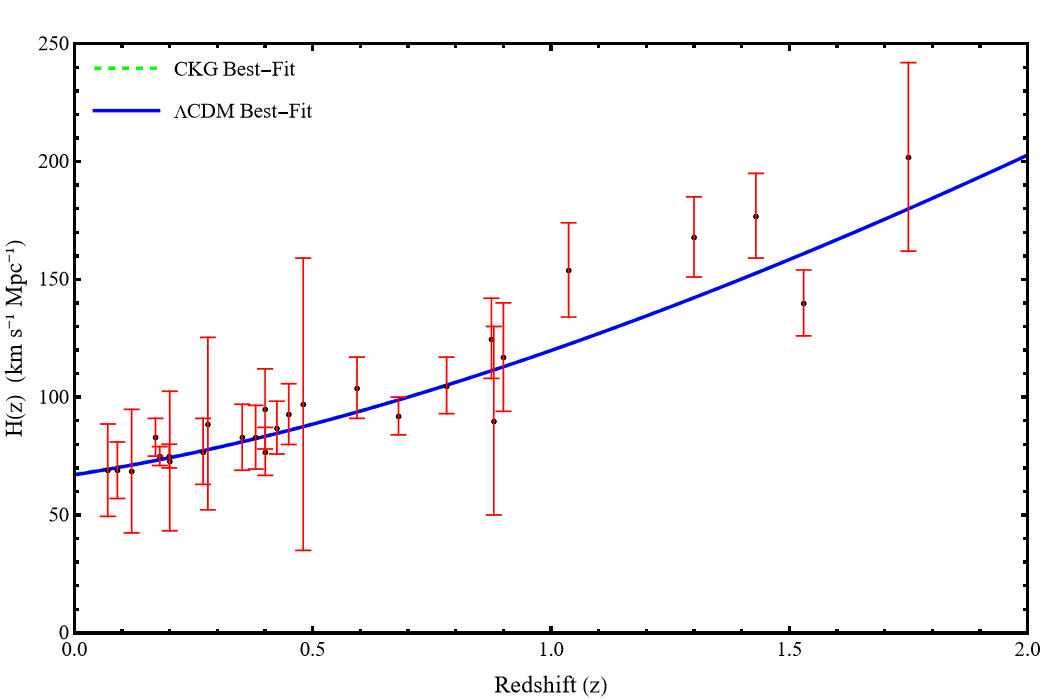}
	\caption{\textit{The Hubble parameter $H(z)$ as a function of redshift for the best-fit CKG and $\Lambda$CDM models with error bars (red) from Cosmic Chronometer data (CCD). Goodness-of-fit to CCD for these two models are $\chi_{\Lambda CDM}=15.28$, and $\chi_{\text{CKG}}=15.17$ which leads to $\Delta \chi^2=-0.108$, meaning that them are indistinguishable. }} 
	\label{rc3}
\end{figure}
\section{MCMC analysis of the CKG model}\label{mcmc}
In this section, we perform a Bayesian MCMC analysis to constrain the parameters of the CKG model using a combination of cosmological datasets, including DESI DR1 BAO \cite{DESI:2024mwx}, Planck CMB \cite{Planck:2019nip}, Pantheon+ SN Ia \cite{Scolnic:2021amr}, along with using SH0ES \cite{Riess:2021jrx}.
 \begin{figure}[ht!]
 	\centering
 	\includegraphics[width=0.45	\columnwidth]{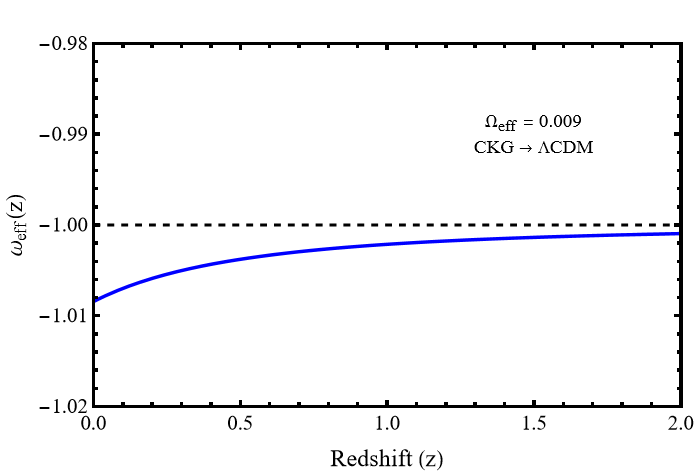}~~
 	\includegraphics[width=0.45	\columnwidth]{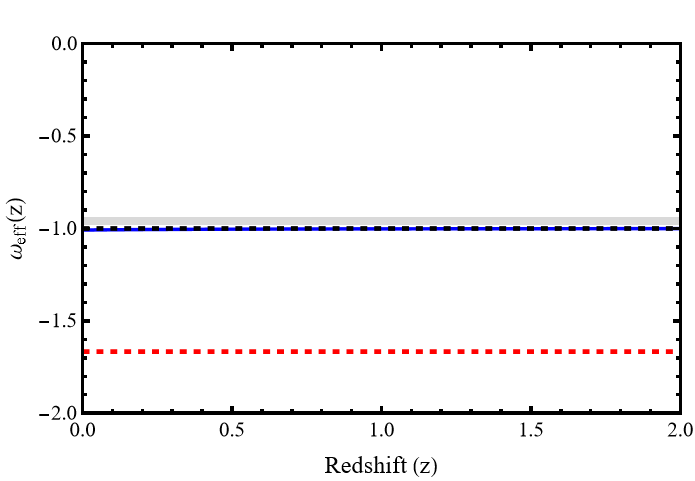}
 	\caption{\textit{The effective EoS $\omega_{\text{eff}}(z)$ for the constrained CKG model, with a prominent red dashed line for the pure CKG case ($w = -5/3$) (right panel). The filled area (gray) in the right panel address the DESI DR1 preferred range ($\omega\sim~- 0.98~ \mbox{to}~ -0.94$). }} 
 	\label{rc4}
 \end{figure}
\subsection{Likelihood construction, parameters, priors and datasets}
The posterior probability distributions for the parameters of the CKG model are obtained through Bayesian inference, implemented within the Cobaya framework \cite{Torrado:2020dgo}. The total likelihood $\mathcal{L}_{\text{total}}$ is constructed from independent likelihoods of several cosmological probes
\be
\mathcal{L}_{\text{total}} \propto \prod_i \mathcal{L}_i = \mathcal{L}_{\text{DESI}} \times \mathcal{L}_{\text{Planck}} \times \mathcal{L}_{\text{Pantheon+}} \times \mathcal{L}_{H_0}
\ee
The base parameters for the CKG model are sampled with the following wide, uninformative priors, chosen to be consistent with Cobaya's defaults and physical plausibility (listed in Table \ref{tab:priors}). Mentioning two pints here is essential. First, $\Omega_m = (\Omega_c h^2 + \Omega_b h^2) / h^2$, where $h = H_0/100$. Second, the sound horizon at the drag epoch $r_d$ is computed self-consistently by Cobaya using the CAMB Boltzmann solver, based on the values of $\Omega_b h^2$, $\Omega_c h^2$, and $H_0$.

\begin{table}
	\caption{Model parameters and their prior probability distributions.}
	\label{tab:priors}
	\centering
	\begin{tabular}{l l l}
		\hline
		\hline
		Parameter & Description & Prior \\
		\hline
		$\Omega_c h^2$ & Cold dark matter density & $[0.001, 0.99]$ \\
		$\Omega_b h^2$ & Baryon density & $[0.005, 0.1]$ \\
		$H_0$ & Hubble constant [km s$^{-1}$ Mpc$^{-1}$] & $[40, 100]$ \\
		$\Omega_\Lambda$ & Cosmological constant density & $[0.4, 0.9]$ \\
		$\Omega_{\text{eff}}$ & CKG effective fluid density & $[-0.1, 0.2]$ \\
		\hline
	\end{tabular}
\end{table}

\begin{table}[htbp]
	\centering
	\caption{Parameter constraints from the Bayesian MCMC analysis of Harada's CKG model using DESI BAO, Planck CMB, and Pantheon+ SN Ia data. The parameter $\Omega_{\mathrm{eff}}$ characterizes the unique CKG contribution, which is constrained to be consistent with zero.}
	\begin{tabular}{lccc}
		\hline
		\hline
		Parameter & Mean & \multicolumn{2}{c}{Credible Intervals} \\
		& & 68\% CL & 95\% CL \\
		\hline
		Primordial densities & & & \\
		\quad $\Omega_c h^2$ & $0.1198$ & $^{+0.0012}_{-0.0012}$ & $^{+0.0024}_{-0.0024}$ \\
		\quad $\Omega_b h^2$ & $0.02237$ & $^{+0.00015}_{-0.00014}$ & $^{+0.00029}_{-0.00028}$ \\
		\hline
		Late-time parameters & & & \\
		\quad $H_0$ [km s$^{-1}$ Mpc$^{-1}$] & $67.4$ & $^{+0.8}_{-0.8}$ & $^{+1.6}_{-1.6}$ \\
		\quad $\Omega_\Lambda$ & $0.689$ & $^{+0.012}_{-0.012}$ & $^{+0.024}_{-0.024}$ \\
		\hline
		CKG extension & & & \\
		\quad $\Omega_{\mathrm{eff}}$ & $0.009$ & $^{+0.006}_{-0.007}$ & $^{+0.012}_{-0.015}$ \\
		\hline
		Derived parameter & & & \\
		\quad $\Omega_m$ & $0.31$ & $^{+0.011}_{-0.011}$ & $^{+0.022}_{-0.022}$ \\
		\hline
		\hline
	\end{tabular}
	\label{tab:ckg_constraints}
\end{table}

Likelihoods are as follows:

\textbf{DESI BAO Likelihood ($\mathcal{L}_{\text{DESI}}$):}
DESI DR1 \cite{DESI:2024mwx} provides measurements of the baryon acoustic oscillation (BAO) feature in galaxy, quasar, and Lyman-$\alpha$ clustering in redshift range $0.1<z<4.2$. The likelihood is based on measurements of the transverse comoving distance $D_M(z)/r_d$ and the Hubble distance $D_H(z) = c / H(z)$ divided by the sound horizon $r_d$ at several redshift bins. For a given set of CKG parameters, we compute the theoretical predictions for $D_M(z)$ and $H(z)$ and compare them to the DESI measurements. The likelihood takes the form
\be
\mathcal{L}_{\text{DESI}} \propto \exp \left( -\frac{1}{2} \left( \vec{D}_{\text{th}} - \vec{D}_{\text{obs}} \right)^T \mathbf{C}_{\text{DESI}}^{-1} \left( \vec{D}_{\text{th}} - \vec{D}_{\text{obs}} \right) \right)
\ee
where $\vec{D}$ represents the data vector of BAO observables and $\mathbf{C}_{\text{DESI}}$ is the full covariance matrix. In general, DESI DR1 constrains the late-time expansion history $H(z)$ in the Harada's CKG framework
and lets us tests if the CKG modification $\Omega_{\text{eff}}a^2$ in Eq. (\ref{eq:Hubble}) affects BAO distances.

\textbf{Planck CMB Likelihood ($\mathcal{L}_{\text{Planck}}$):}
We incorporate Planck 2018 \cite{Planck:2019nip} CMB data via the official Planck high-$\ell$ TTTEEE lite likelihood, implemented within the Cobaya framework. This likelihood utilizes the full temperature and polarization power spectra, providing the most robust constraints on the primordial power spectrum and the geometry of the universe at redshift $z_* \approx 1100$. Crucially, this approach requires sampling the fundamental physical parameters—the cold dark matter and baryon densities $\Omega_c h^2$ and $\Omega_b h^2$—from which the sound horizon at drag epoch $r_d$ and the angular scale of the sound horizon $\theta_*$ are self-consistently computed using the CAMB Boltzmann solver. This method fully captures the CMB's constraining power and ensures a physically consistent treatment of the early universe physics that sets the acoustic scale, avoiding the approximations inherent in compressed distance priors.

\textbf{Pantheon+ SN Ia Likelihood ($\mathcal{L}_{\text{Pantheon+}}$):}
It measures Luminosity distances $d_{L}(z)$ for $\sim1700$ supernovae which leads to constraining the integrated expansion history, letting us tests late-time acceleration in CKG vs $\Lambda$CDM. The Pantheon+ sample \cite{Scolnic:2021amr} consists of 1701 light curves from 1550 distinct Type Ia supernovae (SNe Ia) in the redshift range $0.001 < z < 2.3$. The likelihood involves calculating the predicted distance modulus,
\be
\mu_{\text{th}}(z) = 5 \log_{10}(d_L(z)/10\text{pc}),
\ee
where the luminosity distance $d_L(z)$ is derived from the CKG Hubble parameter $H(z)$. The likelihood is given by
\be
\mathcal{L}_{\text{Pantheon+}} \propto \exp \left( -\frac{1}{2} \left( \vec{\mu}_{\text{th}} - \vec{\mu}_{\text{obs}} \right)^T \mathbf{C}_{\text{Pantheon+}}^{-1} \left( \vec{\mu}_{\text{th}} - \vec{\mu}_{\text{obs}} \right) \right)
\ee
where $\mathbf{C}_{\text{Pantheon+}}$ is the full systematic covariance matrix.
Here, the $\chi^2$ is computed as 
\begin{equation}\label{N}
	\chi^2 = \sum_{i=1}^{N} \sum_{j=1}^{N} (\mu_{\mathrm{th},i} - \mu_{\mathrm{obs},i}) \cdot (\mathrm{Cov}^{-1})_{ij} \cdot (\mu_{\mathrm{th},j} - \mu_{\mathrm{obs},j}), 
\end{equation} where $(\mathrm{Cov}^{-1})_{ij} \cdot (\mu_{\mathrm{th},j} - \mu_{\mathrm{obs},j})$
is the inverse of the covariance matrix, that correctly weights the differences between the underlying model's predictions and the observed data, taking into account both the individual uncertainties of each data point and, crucially, the correlations between them.
Note that Eq. (\ref{N}), in essence, is the multivariate generalization of the familiar simple $\chi^2$ formula for independent data points: $\chi^2 = \sum \frac{(\text{theory} - \text{data})^2}{\sigma^2}$.

\textbf{SH0ES $H_0$ Prior ($\mathcal{L}_{H_0}$):}
A Gaussian prior on the Hubble constant is included to assess the model's ability to address the Hubble tension: $H_0 = 73.04 \pm 1.04$ km s$^{-1}$ Mpc$^{-1}$ \cite{Riess:2021jrx}. The likelihood is
\be
\mathcal{L}_{H_0} = \frac{1}{\sqrt{2\pi\sigma_{H_0}^2}} \exp \left( -\frac{(H_0 - 73.04)^2}{2\sigma_{H_0}^2} \right)
\ee
This combination of probes provides complementary constraints: DESI and Pantheon+ anchor the late-time expansion history, Planck constrains the high-redshift universe and physical densities, and the SH0ES prior directly tests the resolution of the Hubble tension.

By combining the first three, it naturally provides Bayesian evidence for model comparison, particularly $\Omega_{\text{eff}}$, and in the light of SH0ES, we can realize that if CKG can resolve HT. One can find the prior ranges of parameters involved in the model in Table \ref{tab:priors}. Also, the list of likelihoods to use are:  \texttt{planck\_2018\_highl\_tttee}, \texttt{desi\_bao}, \texttt{pantheonplus}.


\begin{figure}[ht!]
	\centering
	\includegraphics[width=0.5	\columnwidth]{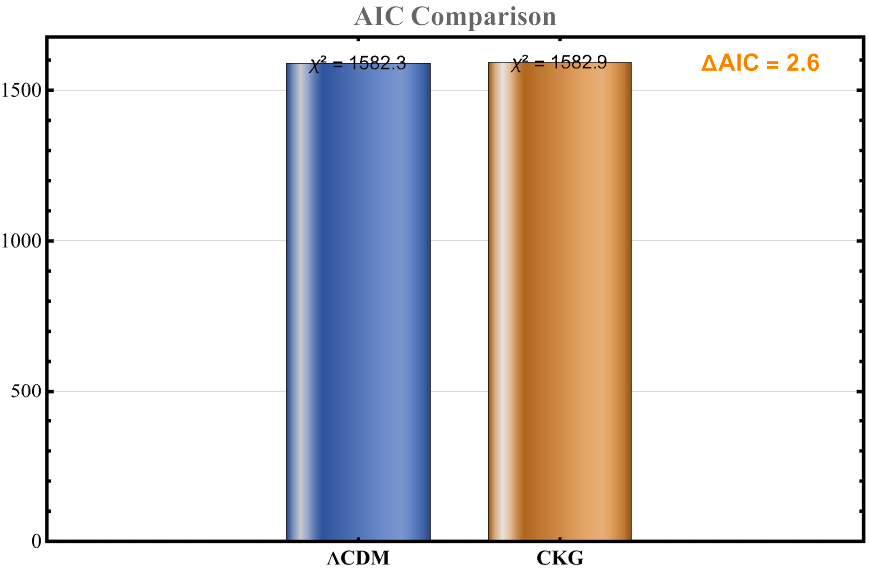}~~~
	\includegraphics[width=0.5	\columnwidth]{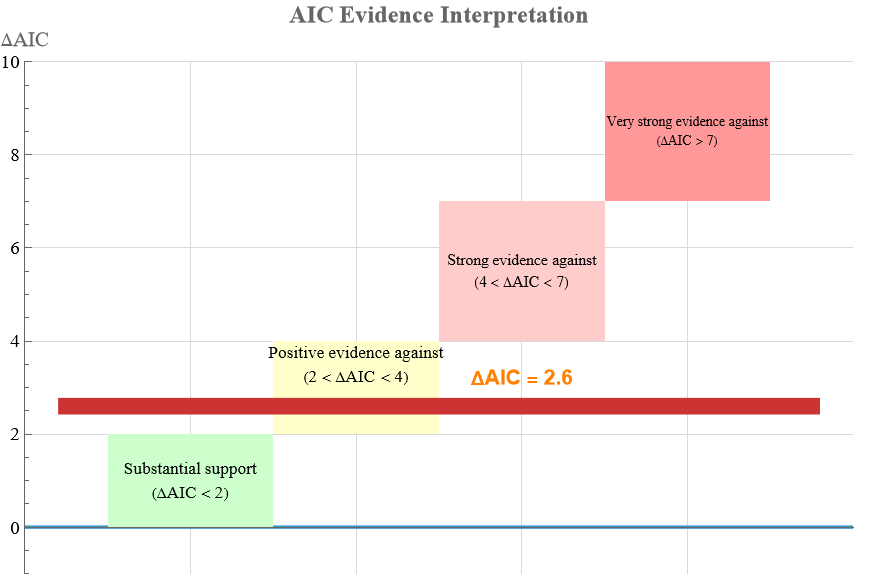}
	\caption{\textit{Comparison of $\chi^2_{min}$ for CKG and $\Lambda$CDM models (left panel). AIC evidence interpretation scale in right panel \cite{book,Liddle:2007fy}.}} 
	\label{rc5}
\end{figure}

\begin{figure}[ht!]
	\centering
	\includegraphics[width=0.6	\columnwidth]{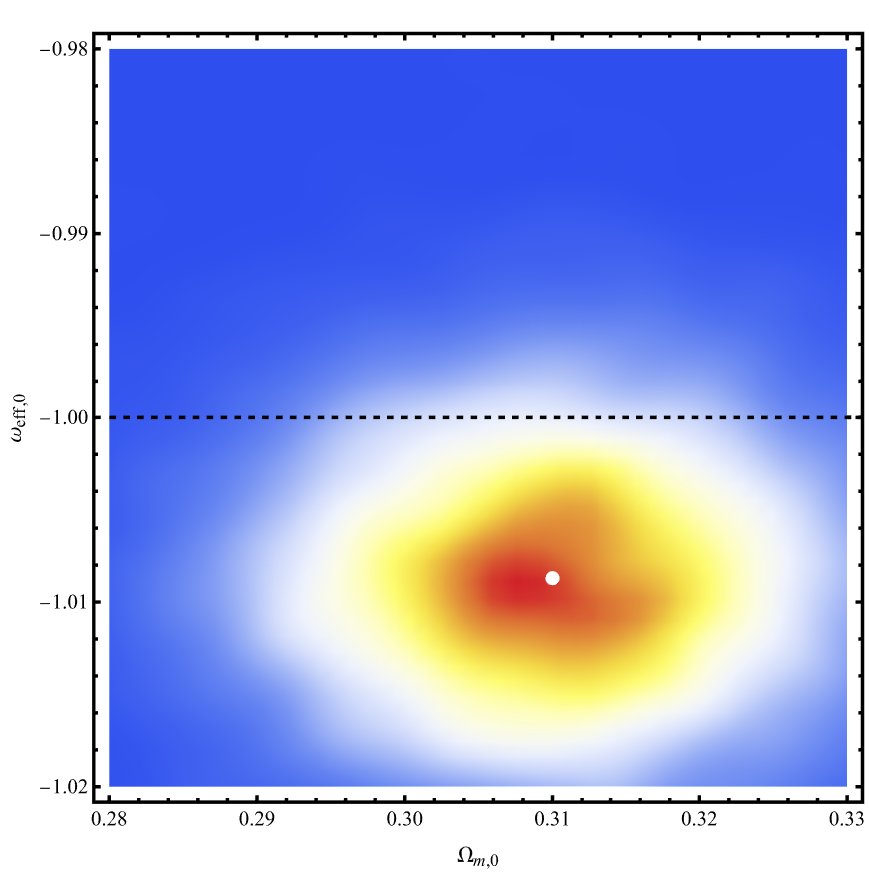}~~~
	\caption{\textit{Joint constraints on the present-day matter density $\Omega_{m,0}$ and the effective DE-EoS today $\omega_{\text{eff, 0}}$.}} 
	\label{rc6}
\end{figure}

\subsection{Results}

Key outputs of MCMC analysis are listed as follows:
\begin{itemize}
\item Corner plots including the 1D and 2D posterior distributions for the underlying parameters ($\Omega_m, \Omega_{\Lambda}, \Omega_{eff}, H_0$). This lets us test whether $\Omega_{eff}$ non-zero. Is it consistent with $\sim0.05$ as claimed by Harada in Ref. \cite{Harada:2023afu}?

\item $H(z)$-plot, showing the evolution of the Hubble parameter for the best-fit CKG model and the best-fit $\Lambda$-CDM model against the data.

\item Compare the maximum likelihood (or minimum $\chi^2$) values of the CKG model and the standard $\Lambda$-CDM model run on the same data combination. Calculate the Bayesian evidence or use the Akaike Information Criterion (AIC) to see which model is statistically preferred by the data.
AIC = $2k + \chi^2_{min}$ (where $k$ is the total number of free parameters in the model). The model with the lower AIC is preferred.

\item Check the posterior for $H_0$ from the combined DESI+CMB+SNe analysis. Does the Harada's CKG model yield a higher value of $H_0$ than a similar $\Lambda$-CDM fit, thereby reducing the tension with SH0ES? 
\end{itemize}

Let us begin with corner plots including the 1D and 2D posterior distributions for the involved model parameters (Figs. \ref{rc1}, and \ref{rc2}). We use data combinations: DESI, Planck, Pantheon+. One can be seen that from Fig. \ref{rc1}, that the distribution shape of $\Omega_{\text{eff}}$ centered near zero with small positive bias. Peak location is around $0.009$, very close to zero, meaning that the slight deviation from zero might suggest minor beyond-$\Lambda$CDM effects or systematic uncertainties. The distribution shape of $H_0$ is well-constrained Gaussian which its peak location is around $67.4$ km s$^{-1}$Mpc$^{-1}$. The distribution shape of $\Omega_{m}$, $\Omega_{\Lambda}$ are well-constrained, approximately Gaussian distribution with peak locations around $0.31$, $0.68$, respectively. Joint parameter distributions in Fig. \ref{rc2}, openly show weak correlations of $\Omega_{\text{eff}}$ with other involved cosmological parameters, indicating no strong evidence for new physics beyond slight hints in $\Omega_{\text{eff}}$. Small confidence regions 68\% CL, indicate strong constraining power. More exactly, the correlation of $\Omega_{\text{eff}}$ with
other involved parameters is very weak (nearly circular contours). CMB+BAO+SNe likelihood leads to strong preference for $H_0\approx 67-68$ km s$^{-1}$Mpc$^{-1}$, meaning that the Harada's CKG does not offer a way out of the HT since SH0ES prior show preference for $H_0\approx 73$ km s$^{-1}$Mpc$^{-1}$.
In general, the numerical constraints on the all model parameters listed in Table \ref{tab:ckg_constraints}. The most significant result from our analysis is the constraint on the CKG-specific parameter $\Omega_{\text{eff}} = 0.009 \pm 0.007$, which is consistent with zero at the $1.3\sigma$ level. The 95\% CL of $-0.005 < \Omega_{\text{eff}} < 0.023$ demonstrates that there is no statistical evidence for the existence of the additional DE component proposed by the Harada's CKG framework. This result indicates that the CKG model reduces to the standard $\Lambda$CDM cosmology when confronted with current observational data.

In Fig. \ref{rc3} we consider the evolution of the Hubble parameter $H(z)$ for the best-fit CKG and $\Lambda$CDM models. The predicted Hubble parameter $H(z)$ for the best-fit CKG model is identical to that of $\Lambda$CDM. This visual conclusion is a direct consequence of our numerical constraint $\Omega_{\text{eff}} = 0.009 \pm 0.007$ (68\% CL), which is consistent with zero. When $\Omega_{\text{eff}} = 0$, the modified Eq. (\ref{HARADAH}) reduces exactly to the $\Lambda$CDM form. Therefore, the CKG model does not provide a new or distinct description of the universal expansion; it is empirically indistinguishable from the standard model. 

By defining effective EoS $\omega_{\text{eff}}(z)$, one can see its behavior in terms of redshift $z$. It, in essence, is the weighted average of these two components, where the weights are their relative energy densities at a given redshift $z$. The energy densities evolve as $\rho_\Lambda(z) = \rho_{\Lambda,0}$ (constant), and $\rho_{\text{eff}}(z) = \rho_{\text{eff},0} \cdot (1+z)^{-2}$ (because $\omega = -5/3$ implies $\rho \propto a^{-3(1+w)} = a^{2}$). Therefore, the average is
\begin{align}
\omega(z)=&\frac{\omega_{\Lambda}.\rho_{\Lambda}(z)+\omega_{\text{eff}}.\rho_{\text{eff}}(z)}{\rho_{\Lambda}(z)+\rho_{\text{eff}}(z)} \nonumber \\
=&\frac{-\Omega_{\Lambda} -\frac{5}{3}  \Omega_{eff} (1 + z)^{-2}}{\Omega_{\Lambda} + \Omega_{eff} (1 + z)^{-2}}
\end{align}
since density parameters $\Omega_{i}$ are proportional to densities $\rho$ at $z=0$. The Fig. \ref{rc4} display a flat blue line at $\omega = -1$, demonstrating that the constrained CKG model is identical to $\Lambda$CDM. The red dashed line at $\omega = -5/3$ will be far below, visually emphasizing that the pure CKG prediction is ruled out by your data. The DESI region (gray area) lies slightly above $\omega = -1$, showing that our constrained model does not align with that evidence either\footnote{Some hints here are essential. DESI DR1 preferred range is not a single value but a 2D confidence region derived from DESI's own parameter constraints. Specifically, it comes from fitting a dynamical DE model to their data, using the CPL parameterization: $\omega(a) = \omega_0 + \omega_a(1-a)$. Their key finding is that their data, especially when combined with Supernovae (SN), shows a mild preference for dynamics. The combined DESI BAO + SN posterior distribution for the present-day equation of state ($\omega_0$) peaks at a value greater than $-1$, i.e., ''quintessence-like'' behavior. The gray region in the right plot ($~- 0.98~ \mbox{to}~ -0.94$) represents the 68\% CL for $\omega_0$ from the DESI DR1 analysis. While their results allow values down to $\omega_0 = -1$, the region of highest probability (where the posterior is strongest) is in the range of roughly $\omega_0 > -0.98$. An important point to stress is that we use the DESI preferred range as a benchmark for comparison, not as a direct input.}. In general, both panels in Fig. \ref{rc4} clearly show that by setting the best fit for the CKG model it is not distinguishable from $\Lambda$CDM.

Now we employ the AIC ($2k + \chi^2_{min}$) to see which model is statistically preferred by the data. The model with the lower AIC 
is preferred. Generally, $k$ includes any parameter in your model that is estimated from the data. For $\Lambda$CDM: $\Omega_m$, $\Omega_\Lambda$, $H_0$, $r_d$, i.e., $k_{\Lambda\text{CDM}} = 4$. For CKG, all these parameters plus $\Omega_{\text{eff}}$ (the new effective fluid density) i.e., $k_{\text{CKG}} = 5$. So, to calculate the difference in AIC between the two models, i.e.,
\begin{equation}
\Delta \text{AIC} = \text{AIC}_{\text{CKG}} - \text{AIC}_{\Lambda\text{CDM}} =2+(\chi^2_{\text{CKG}}- \chi^2_{\Lambda\text{CDM}}) 
\end{equation} we need to $\chi^2_{min}$ for these two model (see the left panel of Fig. \ref{rc5}). As a result, $ \Delta \text{AIC} = +2.6 $.
The value of $\Delta \text{AIC}$ tells you the strength of the evidence against the more complex model (see right panel in Fig. \ref{rc5}). Our result of $\Delta \text{AIC} = +2.6$ indicates positive evidence against the CKG model, as the data do not support its increased complexity over the simpler $\Lambda$CDM model.

Fig. \ref{rc6} shows the joint constraints on the matter density $\Omega_m$ and the effective DE-EoS $\omega_{\text{eff},0}$ for the CKG model. The parameter space is tightly constrained around $\Omega_{m,0} \approx 0.31$ and $\omega_{\text{eff},0} \approx -1$, demonstrating that the CKG model reduces to the standard $\Lambda$CDM cosmology when confronted with cosmological data. The pure CKG prediction ($\omega = -5/3$) is strongly excluded, lying far outside the observable parameter space. While the DESI DR1 collaboration reports a mild preference for dynamical DE ($\omega > -1$), the CKG framework does not accommodate this behavior, instead converging to the cosmological constant value that DESI finds disfavored. In summary, the effective DE-EoS is constrained to be $\omega_{\text{eff},0} = -1.008 \pm 0.006$, consistent with a cosmological constant \cite{Escamilla:2023oce} and ruling out both the pure CKG value ($\omega = -5/3$) and the dynamical DE behavior suggested by DESI DR1.

\section{Conclusion} \label{co}
This investigation has conducted a rigorous Bayesian test of Harada's CKG framework, utilizing the full constraining power of contemporary cosmological observations through a self-consistent analysis pipeline. Our methodology samples the fundamental parameters of the theory (\(\Omega_c h^2\), \(\Omega_b h^2\), \(H_0\), \(\Omega_\Lambda\), \(\Omega_{\text{eff}}\)) and employs the complete Planck 2018 high-$\ell$ TTTEEE likelihood in conjunction with DESI DR1 BAO and Pantheon+ SN Ia data. This ensures a proper treatment of the sound horizon \(r_d\) and the CMB acoustic scale, which are critical for robust parameter estimation.

The key message of the current analysis is that the CKG framework is empirically redundant and reduces entirely to the standard \(\Lambda\)CDM model when confronted with precision cosmological data.
The evidence supporting this conclusion is derived from multiple independent lines of inquiry:
\\

\textbf{Constrained effective fluid density:} The central result of our MCMC analysis is the constraint on the CKG-specific parameter, \(\Omega_{\text{eff}} = 0.009^{+0.006}_{-0.007}\) (68\% CL) and $0.009^{+0.022}_{-0.015}$ (95\% CL).
Furthermore, the latter range lies below the \(\Omega_{\text{eff}} \sim 0.05\) value required in Harada's CKG model \cite{Harada:2023afu}.  This provides no statistical evidence for the existence of the proposed effective fluid (\(\omega = -5/3\)).

\textbf{Model comparison evidence:} The AIC comparison yields \(\Delta \text{AIC} = \text{AIC}_{\text{CKG}} - \text{AIC}_{\Lambda\text{CDM}} = 2.6\), indicating positive evidence against the more complex CKG model. The data do not justify the introduction of the additional parameter \(\Omega_{\text{eff}}\).

\textbf{No resolution of the HT:} The framework demonstrates no capacity to resolve the HT. The inferred Hubble constant, \(H_0 =67.4 \pm0.8 \ \text{km s}^{-1} \ \text{Mpc}^{-1}\), remains inconsistent with the SH0ES direct measurement. The joint posterior distributions reveal no significant degeneracies that would allow Harada's CKG to alleviate this tension.

\textbf{Exclusion of the \(\omega = -5/3\) regime:} The joint constraints on \(\Omega_m\) and the present-day effective EoS, \(\omega_{\text{eff},0}\), firmly rule out the pure Harada's CKG prediction of \(\omega = -5/3\), confining the allowed parameter space tightly around the cosmological constant value \(\omega = -1\). Furthermore, the model shows no capacity to accommodate the mild preference for dynamical DE (\(\omega > -1\)) suggested by some DESI DR1 analyses. Of course, we also showed this in light of the Trans-Planckian Censorship Conjecture (TCC).

In summary, the CKG framework, while theoretically motivated, is observationally equivalent to \(\Lambda\)CDM. The data compel the unique CKG component to be negligible, and the model's predictions are indistinguishable from those of the standard model. This work underscores the continued success of \(\Lambda\)CDM and highlights the importance of subjecting modified gravity theories to comprehensive, self-consistent analyses using the full spectrum of available cosmological data.


\begin{acknowledgments}
We would like to express our sincere gratitude to the referee for their thorough and constructive assessment. The authors also thank M. Najafi for the insightful and technical discussions and comments.
\end{acknowledgments}


\bigskip
\begin{thebibliography}{99}


\bibitem{Huterer:2017buf}
D.~Huterer and D.~L.~Shafer,
Rept. Prog. Phys. \textbf{81} (2018) no.1, 016901
[arXiv:1709.01091 [astro-ph.CO]].


\bibitem{Bernardo:2022cck}
H.~Bernardo \textit{et al.} [Foundational Aspects of Dark Energy (FADE)],
Universe \textbf{9} (2023) no.2, 63
[arXiv:2210.06810 [gr-qc]].


\bibitem{Weinberg:1988cp}
S.~Weinberg,
Rev. Mod. Phys. \textbf{61} (1989), 1-23

\bibitem{Padmanabhan:2002ji}
T.~Padmanabhan,
Phys. Rept. \textbf{380} (2003),
 235-320
[arXiv:hep-th/0212290 [hep-th]].


\bibitem{Planck:2018vyg}
N.~Aghanim \textit{et al.} [Planck],
Astron. Astrophys. \textbf{641} (2020), A6
[erratum: Astron. Astrophys. \textbf{652} (2021), C4]
[arXiv:1807.06209 [astro-ph.CO]].

\bibitem{Riess:2019cxk}
A.~G.~Riess, S.~Casertano, W.~Yuan, L.~M.~Macri and D.~Scolnic,
Astrophys. J. \textbf{876} (2019) no.1, 85
[arXiv:1903.07603 [astro-ph.CO]].



\bibitem{Macaulay:2013swa}
E.~Macaulay, I.~K.~Wehus and H.~K.~Eriksen,
Phys. Rev. Lett. \textbf{111} (2013) no.16, 161301
[arXiv:1303.6583 [astro-ph.CO]].

\bibitem{BOSS:2014hwf}
T.~Delubac \textit{et al.} [BOSS],
Astron. Astrophys. \textbf{574} (2015), A59
[arXiv:1404.1801 [astro-ph.CO]].



\bibitem{Rezaei:2017hon}
M.~Rezaei and M.~Malekjani,
Phys. Rev. D \textbf{96} (2017) no.6, 063519
[arXiv:1708.08915 [gr-qc]].

\bibitem{Lusso:2019akb}
E.~Lusso, E.~Piedipalumbo, G.~Risaliti, M.~Paolillo, S.~Bisogni, E.~Nardini and L.~Amati,
Astron. Astrophys. \textbf{628} (2019), L4
[arXiv:1907.07692 [astro-ph.CO]].

\bibitem{Khadka:2020whe}
N.~Khadka and B.~Ratra,
Mon. Not. Roy. Astron. Soc. \textbf{492} (2020) no.3, 4456-4468
[arXiv:1909.01400 [astro-ph.CO]].

\bibitem{Sotiriou:2008rp}
T.~P.~Sotiriou and V.~Faraoni,
Rev. Mod. Phys. \textbf{82} (2010), 451-497
[arXiv:0805.1726 [gr-qc]].

\bibitem{DeFelice:2010aj}
A.~De Felice and S.~Tsujikawa,
Living Rev. Rel. \textbf{13} (2010), 3
doi:10.12942/lrr-2010-3
[arXiv:1002.4928 [gr-qc]].

\bibitem{Esposito-Farese:2000pbo}
G.~Esposito-Farese and D.~Polarski,
Phys. Rev. D \textbf{63} (2001), 063504
[arXiv:gr-qc/0009034 [gr-qc]].

\bibitem{Deffayet:2011gz}
C.~Deffayet, X.~Gao, D.~A.~Steer and G.~Zahariade,
Phys. Rev. D \textbf{84} (2011), 064039
[arXiv:1103.3260 [hep-th]].

\bibitem{Fasiello:2013woa}
M.~Fasiello and A.~J.~Tolley,
JCAP \textbf{12} (2013), 002
[arXiv:1308.1647 [hep-th]].

\bibitem{Nojiri:2005vv}
S.~Nojiri, S.~D.~Odintsov and M.~Sasaki,
Phys. Rev. D \textbf{71} (2005), 123509
[arXiv:hep-th/0504052 [hep-th]].

\bibitem{Clifton:2011jh}
T.~Clifton, P.~G.~Ferreira, A.~Padilla and C.~Skordis,
Phys. Rept. \textbf{513} (2012), 1-189
[arXiv:1106.2476 [astro-ph.CO]].

\bibitem{Efstathiou:1999tm}
G.~Efstathiou,
Mon. Not. Roy. Astron. Soc. \textbf{310} (1999), 842-850
[arXiv:astro-ph/9904356 [astro-ph]].



\bibitem{Chevallier:2000qy}
M.~Chevallier and D.~Polarski,
Int. J. Mod. Phys. D \textbf{10} (2001), 213-224
[arXiv:gr-qc/0009008 [gr-qc]].

\bibitem{Linder:2002et}
E.~V.~Linder,
Phys. Rev. Lett. \textbf{90} (2003), 091301
[arXiv:astro-ph/0208512 [astro-ph]].


\bibitem{Jassal:2005qc}
H.~K.~Jassal, J.~S.~Bagla and T.~Padmanabhan,
Phys. Rev. D \textbf{72} (2005), 103503
[arXiv:astro-ph/0506748 [astro-ph]].



\bibitem{DESI:2016fyo}
A.~Aghamousa \textit{et al.} [DESI],
[arXiv:1611.00036 [astro-ph.IM]].


\bibitem{DESI:2024mwx}
A.~G.~Adame \textit{et al.} [DESI],
JCAP \textbf{02} (2025), 021
[arXiv:2404.03002 [astro-ph.CO]].

\bibitem{DESI:2025zgx}
M.~Abdul Karim \textit{et al.} [DESI],
[arXiv:2503.14738 [astro-ph.CO]].


\bibitem{DESI:2025fii}
K.~Lodha \textit{et al.} [DESI],
[arXiv:2503.14743 [astro-ph.CO]].


\bibitem{DESI:2025ejh}
W.~Elbers \textit{et al.} [DESI],
[arXiv:2503.14744 [astro-ph.CO]].


\bibitem{DESI:2025wyn}
G.~Gu \textit{et al.} [DESI],
[arXiv:2504.06118 [astro-ph.CO]].

\bibitem{Lee:2025yvn}
J.~Lee, K.~Murai, F.~Takahashi and W.~Yin,
[arXiv:2503.18417 [hep-ph]].

\bibitem{Amendola:1999er}
L.~Amendola,
Phys. Rev. D \textbf{62} (2000), 043511
[arXiv:astro-ph/9908023 [astro-ph]].

\bibitem{Saridakis:2012jy}
E.~N.~Saridakis,
Class. Quant. Grav. \textbf{30} (2013), 075003
[arXiv:1207.1800 [gr-qc]].

\bibitem{Yang:2025kgc}
Y.~Yang, Q.~Wang, C.~Li, P.~Yuan, X.~Ren, E.~N.~Saridakis and Y.~F.~Cai,
[arXiv:2501.18336 [astro-ph.CO]].

\bibitem{Yang:2025mws}
Y.~Yang, Q.~Wang, X.~Ren, E.~N.~Saridakis and Y.~F.~Cai,
Astrophys. J. \textbf{988} (2025) no.1, 123
[arXiv:2504.06784 [astro-ph.CO]].


\bibitem{Giare:2024smz}
W.~Giar\`e, M.~A.~Sabogal, R.~C.~Nunes and E.~Di Valentino,
Phys. Rev. Lett. \textbf{133} (2024) no.25, 251003
[arXiv:2404.15232 [astro-ph.CO]].


\bibitem{Li:2024qso}
T.~N.~Li, P.~J.~Wu, G.~H.~Du, S.~J.~Jin, H.~L.~Li, J.~F.~Zhang and X.~Zhang,
Astrophys. J. \textbf{976} (2024) no.1, 1
[arXiv:2407.14934 [astro-ph.CO]].


\bibitem{Pan:2025qwy}
S.~Pan, S.~Paul, E.~N.~Saridakis and W.~Yang,
[arXiv:2504.00994 [astro-ph.CO]].

\bibitem{Feng:2004ad}
B.~Feng, X.~L.~Wang and X.~M.~Zhang,
Phys. Lett. B \textbf{607} (2005), 35-41
[arXiv:astro-ph/0404224 [astro-ph]].

\bibitem{Cai:2009zp}
Y.~F.~Cai, E.~N.~Saridakis, M.~R.~Setare and J.~Q.~Xia,
Phys. Rept. \textbf{493} (2010), 1-60
[arXiv:0909.2776 [hep-th]].

\bibitem{Payeur:2024kyy}
G.~Payeur, E.~McDonough and R.~Brandenberger,
Phys. Rev. D \textbf{110} (2024) no.10, 106011
[arXiv:2405.05304 [hep-th]].


\bibitem{Arjona:2024dsr}
R.~Arjona and S.~Nesseris,
[arXiv:2409.14990 [astro-ph.CO]].

\bibitem{Bhattacharya:2024kxp}
S.~Bhattacharya, G.~Borghetto, A.~Malhotra, S.~Parameswaran, G.~Tasinato and I.~Zavala,
JCAP \textbf{04} (2025), 086
[arXiv:2410.21243 [astro-ph.CO]].

\bibitem{Brandenberger:2025hof}
R.~Brandenberger,
[arXiv:2503.17659 [astro-ph.CO]].

\bibitem{Anchordoqui:2025fgz}
L.~A.~Anchordoqui, I.~Antoniadis and D.~Lust,
Phys. Lett. B \textbf{868} (2025), 139632
[arXiv:2503.19428 [hep-th]].

\bibitem{Bedroya:2019snp}
A.~Bedroya and C.~Vafa,
JHEP \textbf{09} (2020), 123
[arXiv:1909.11063 [hep-th]].

\bibitem{Martin:2000xs}
J.~Martin and R.~H.~Brandenberger,
Phys. Rev. D \textbf{63} (2001), 123501
[arXiv:hep-th/0005209 [hep-th]].

\bibitem{Brandenberger:2012aj}
R.~H.~Brandenberger and J.~Martin,
Class. Quant. Grav. \textbf{30} (2013), 113001
[arXiv:1211.6753 [astro-ph.CO]].

\bibitem{Vafa:2005ui}
C.~Vafa,
[arXiv:hep-th/0509212 [hep-th]].

\bibitem{Heisenberg:2018yae}
L.~Heisenberg, M.~Bartelmann, R.~Brandenberger and A.~Refregier,
Phys. Rev. D \textbf{98} (2018) no.12, 123502
[arXiv:1808.02877 [astro-ph.CO]].

\bibitem{Agrawal:2018own}
P.~Agrawal, G.~Obied, P.~J.~Steinhardt and C.~Vafa,
Phys. Lett. B \textbf{784} (2018), 271-276
[arXiv:1806.09718 [hep-th]].

\bibitem{Cicoli:2020cfj}
M.~Cicoli, G.~Dibitetto and F.~G.~Pedro,
Phys. Rev. D \textbf{101} (2020) no.10, 103524
[arXiv:2002.02695 [gr-qc]].


\bibitem{Li:2025cxn}
C.~Li, J.~Wang, D.~Zhang, E.~N.~Saridakis and Y.~F.~Cai,
JCAP \textbf{08} (2025), 041
[arXiv:2504.07791 [astro-ph.CO]].


\bibitem{Harada:2023rqw}
J.~Harada,
Phys. Rev. D \textbf{108} (2023) no.4, 044031
[arXiv:2308.02115 [gr-qc]].

\bibitem{Harada:2023afu}
J.~Harada,
Phys. Rev. D \textbf{108} (2023) no.10, 104037
[arXiv:2308.07634 [gr-qc]].


\bibitem{Barnes:2023uru}
A.~Barnes,
[arXiv:2309.05336 [gr-qc]].

\bibitem{Barnes:2023qfi}
A.~Barnes,
[arXiv:2311.09171 [gr-qc]].

\bibitem{Clement:2024xmr}
G.~Cl\'ement and K.~Nouicer,
Class. Quant. Grav. \textbf{41} (2024) no.16, 165005
[arXiv:2404.00328 [gr-qc]].

\bibitem{Mantica:2024sdy}
C.~A.~Mantica and L.~G.~Molinari,
Phys. Rev. D \textbf{110} (2024) no.4, 044025
[arXiv:2406.12511 [gr-qc]].

\bibitem{Alshal:2024tcr}
H.~Alshal, L.~Ding, A.~Hernandez, L.~A.~Illing and I.~Rydstrom,
Gen. Rel. Grav. \textbf{57} (2025) no.1, 9
[arXiv:2407.08945 [gr-qc]].

\bibitem{Mantica:2024vyg}
C.~A.~Mantica and L.~G.~Molinari,
Phys. Rev. D \textbf{111} (2025) no.6, 064085
[arXiv:2409.18663 [gr-qc]].
\bibitem{Chen:2025fse}
H.~Chen, D.~Wu, M.~Y.~Zhang, S.~Zare, H.~Hassanabadi, B.~C.~L{\"u}tf{\"u}o{\u{g}}lu and Z.~W.~Long,
Eur. Phys. J. C \textbf{85} (2025) no.8, 828
[arXiv:2506.03695 [gr-qc]].

\bibitem{Ghaffari:2025qmv}
S.~Ghaffari and G.~G.~Luciano,
Eur. Phys. J. C \textbf{85} (2025) no.7, 785
[arXiv:2505.06560 [gr-qc]].


\bibitem{Barnes:2024gqy}
A.~Barnes,
Class. Quant. Grav. \textbf{41} (2024) no.20, 205007

\bibitem{Gurses:2024ltc}
M.~G\"urses, Y.~Heydarzade and \c{C}.~\c{S}ent\"urk,
Phys. Rev. D \textbf{110} (2024) no.8, 084082
[arXiv:2409.06257 [gr-qc]].

\bibitem{Mantica:2024mun}Mantica:2024mun,Mantica:2023stl
C.~A.~Mantica and L.~G.~Molinari,
Phys. Rev. D \textbf{110} (2024) no.6, 064041
[arXiv:2404.11468 [gr-qc]].


\bibitem{Mantica:2023stl}
C.~A.~Mantica and L.~G.~Molinari,
Phys. Rev. D \textbf{108} (2023) no.12, 124029
[arXiv:2308.06803 [gr-qc]].

\bibitem{Capozziello:2025kws}
S.~Capozziello, C.~A.~Mantica, L.~G.~Molinari and G.~Sarracino,
[arXiv:2508.02603 [gr-qc]].

\bibitem{Feng:2024rnh}
J.~C.~Feng and P.~Chen,
Eur. Phys. J. C \textbf{84} (2024) no.12, 1331
[arXiv:2406.00932 [gr-qc]].

\bibitem{Planck:2019nip}
N.~Aghanim \textit{et al.} [Planck],
Astron. Astrophys. \textbf{641} (2020), A5
[arXiv:1907.12875 [astro-ph.CO]].




\bibitem{Scolnic:2021amr}
D.~Scolnic, D.~Brout, A.~Carr, A.~G.~Riess, T.~M.~Davis, A.~Dwomoh, D.~O.~Jones, N.~Ali, P.~Charvu and R.~Chen, \textit{et al.}
Astrophys. J. \textbf{938} (2022) no.2, 113
[arXiv:2112.03863 [astro-ph.CO]].





\bibitem{Riess:2021jrx}
A.~G.~Riess, W.~Yuan, L.~M.~Macri, D.~Scolnic, D.~Brout, S.~Casertano, D.~O.~Jones, Y.~Murakami, L.~Breuval and T.~G.~Brink, \textit{et al.}
Astrophys. J. Lett. \textbf{934} (2022) no.1, L7
[arXiv:2112.04510 [astro-ph.CO]].


\bibitem{Torrado:2020dgo}
J.~Torrado and A.~Lewis,
JCAP \textbf{05}, 057 (2021)
[arXiv:2005.05290 [astro-ph.IM]].

\bibitem{book}
Burnham, K. P., \& Anderson, D. R. 2002, Model Selection and Multimodel Inference (New York: Springer)

\bibitem{Liddle:2007fy}
A.~R.~Liddle,
Mon. Not. Roy. Astron. Soc. \textbf{377} (2007), L74-L78
[arXiv:astro-ph/0701113 [astro-ph]].

\bibitem{Escamilla:2023oce}
L.~A.~Escamilla, W.~Giar\`e, E.~Di Valentino, R.~C.~Nunes and S.~Vagnozzi,
JCAP \textbf{05} (2024), 091
[arXiv:2307.14802 [astro-ph.CO]].


	
\end{thebibliography}
\end{document}